\documentclass[11pt]{iopart}
\usepackage{iopams}
\usepackage{epsfig,cite}
\newcommand{\be}{\begin{equation}}
\newcommand{\ee}{\end{equation}}
\newcommand{\bd}{\begin{displaymath}}
\newcommand{\ed}{\end{displaymath}}
\newcommand{\BE}{\begin{eqnarray}}
\newcommand{\EE}{\end{eqnarray}}

\newcommand{\bra}{\left\langle}
\newcommand{\ket}{\right\rangle}

\newcommand{\id}{{\rm 1\!\! I}}

\newcommand{\ba}{\ensuremath{\mathbf{a}}}

\newcommand{\bx}{\ensuremath{\mathbf{x}}}

\newcommand{\avg}[1]{\left\langle{#1}\right\rangle}

\begin{document}

\title{Minority games, evolving capitals and replicator dynamics}

\author{Tobias Galla${}^{1,2,4}$\footnote{TG's permanent address is at the University of Manchester.} and Yi-Cheng Zhang${}^{1,3}$}

\address{
${}^{1}$ Research Centre for Complex Systems Science, University of Shanghai for Science and Technology, Shanghai 200093, People's Republic of China \\
${}^{2}$ The University of Manchester, School of Physics and Astronomy, Manchester M13 9PL, United Kingdom\\
${}^{3}$ Department of Physics, University of Fribourg, Chemin du Mus\'ee, Fribourg CH-1700, Switzerland\\
${}^{4}$ The Abdus Salam International Centre for Theoretical Physics, Strada Costiera 11, and CNR-INFM Trieste-SISSA Unit, Via Beirut 2-4, 34014 Trieste, Italy}

\begin{abstract}
We discuss a simple version of the Minority Game (MG) in which agents hold
only one strategy each, but in which their capitals evolve dynamically
according to their success and in which the total trading volume varies in
time accordingly. This feature is known to be crucial for MGs to reproduce
stylised facts of real market data. The stationary states and phase diagram of the
model can be computed, and we show that the ergodicity breaking phase transition common for MGs, and marked by a divergence of the integrated response is present also in this simplified model. An analogous majority game turns out to be relatively void of interesting features, and the total capital is found to diverge in time. Introducing a restraining force leads to a model akin to replicator dynamics of evolutionary game theory, and we demonstrate that here a different type of phase transition is observed. Finally we briefly discuss the
relation of this model with one strategy per player to more sophisticated Minority Games with dynamical capitals and several trading strategies per
agent.

\end{abstract}


\ead{\tt Tobias.Galla@manchester.ac.uk, yi-cheng.zhang@unifr.ch}

\section{Introduction}
The statistical mechanics of the Minority Game (MG) in its original
version is now well understood. Since its conception \cite{challetzhang} sophisticated tools
from equilibrium and non-equilibrium statistical physics have been
applied to this simple model of inductive adaptation, and the phase
diagram and behaviour in the ergodic regime have been thoroughly
studied \cite{book1,book2}, see also \cite{rev}. The model is now essentially considered to be solved and its
key observables can be calculated either fully exactly or at least in
good approximation, even though some challenging mathematical issues
are still not fully resolved and deserve further attention \cite{book2}. These
relate mainly to the development and refinement of dynamical theories
to address MGs and related market models with so-called real histories
and to the behaviour of the model in the non-ergodic, turbulent
regime, where satisfactory analytical solutions are as yet still
awaited.

With the main analytical tools to study MGs now well developed and in
place one possible future direction appears to be the extension of
present basic versions of the MG towards more realism as a market
model while at the same time maintaining the analytical tractability
of the original setup. The goal is here to study versions of the MG
which are able to capture features of real market data, so-called
stylised facts, such as non-Gaussian return distributions or
volatility clustering.

The main drawback of the original MG, preventing the emergence of
stylised facts, rests in the assumption all agents contribute equally to the total
outcome, they trade one unit of the underlying asset at any time step
of the dynamics, regardless of their success or otherwise in the
trading. This results in a constant trading volume in time (given simply by
the number of players). It has been demonstrated in that a
modulation of the overall trading volume in the MG can suffice to
obtain stylised facts. Two alternative approaches have been proposed
to allow the trading value to be dynamic. In so-called grand-canonical
MGs (GCMG) \cite{gcmg} agents are endowed with a null strategy and can
decide not to trade at a given time step if they do not consider
trading profitable. Hence the total trading value varies in time. The
GCMG has been found to exhibit volatility clustering and non-Gaussian
return distributions in a critical region around its phase
transition. On the other hand GCMGS are tractable fully analytically
and their phase diagrams and order parameters in the ergodic regime
have been computed. A second option is to endow each agents in the MG with an individual capital, which may vary in time. Such MGs have been introduced in \cite{dyncap}, here agents are equipped with
multiple trading strategies and a dynamical capital, which evolves
according to their success in trading. It is then assumed that any
agent trades a constant fraction of his capital at any time-step, so
that the overall trading volume again varies over time. The analysis
of \cite{dyncap}, based on numerical simulations, shows that such an MG with
dynamical capitals and multiple strategies per player again exhibits
stylised facts, provided model parameters are tuned suitably.

The aim of the present paper is to study the most simple MG with
evolving capitals, namely one in which every agent only holds only
{\em one} trading strategy, but in which at the same time the agent's
capital and her trading volume changes in time according to her
success in the past. The traders are thus zero-intelligence agents, or
simple automata following a strategy pre-set at the beginning of the
game, and with no power to take any decisions or to alter their
behaviour during the course of the game. Allowing for a dynamics of
the weights of each trading strategy here introduces an evolutionary
mechanism, selecting more successful strategies over time. We here
address two different types of dynamics. The first part of the paper
is concerned with a minority game with dynamical capitals. To obtain
non-trivial dynamics we consider a setup in which agents with
dynamical capitals (so-called `speculators') are coupled to a
background of `producers', that is traders who trade constant volumes,
even if in the long-run they run losses. The addition of producers to a population of speculators is common also in GCMGs \cite{book1,book2}. Secondly, we then focus on
Majority Games within a similar setup of dynamical capitals (here there is no need to consider producers), and then
finally we show that a close relation between replicator dynamics in evolutionary
game theory \cite{Book4} and the minority and majority games with dynamical capitals can be drawn. The payoff structure of the minority/majority game here leads to a replicator model with Hebbian/anti-Hebbian couplings. Extending the work of \cite{myhebb} we study these models by means of dynamical approaches of disordered systems
theory, and derive the properties of their stationary fixed point
states as well as their phase diagrams. We find that the common MG-phase
transition between an efficient and a non-efficient phase persists in the setup of the MG with evolving capitals and only one trading strategy per
player. The majority game with dynamical capitals is relatively featureless, but a different type of transition is found in the corresponding replicator dynamics. Ergodicity is broken in model the MG with evolving capitals and in the majority game replicator dynamics, but no divergence of the dynamic susceptibility is found in the latter case. While we focus
on the statistical mechanics properties of both models, we also
briefly address the emergence of stylised facts in this simple MG with
dynamical capitals, and leave a more detailed discussion to future
work \cite{matteoandi}.
   
\section{Minority Game with dynamical capitals}
\subsection{Model}
We consider a system of $N$ agents, labelled by $i=1,\dots,N$. At each round
of the game each agent makes a bid that can be either positive or negative,
and which depends on a common external signal, which is available to all
agents In the simple setup of one `trading strategy' per agent, the sign of
agent $i$'s action is determined by his strategy vector $\ba_i$, which maps
the value of an external piece of information onto the
binary set $\{-1,1\}$. The common random external information is here
represented by an integer number $\mu(t)\in\{1,\dots,P\}$ and is drawn from a
flat distribution over this set at each time step without temporal
correlations.  Thus, each trading strategy $\ba_i$, $i=1,\dots,N$ can be
understood as a vector $\ba_i=(a_i^1,\dots,a_i^{P})\in\{-1,1\}^{P}$, mapping
information $\mu$ onto a trading decision of sign $a_i^\mu$. We will assume
the amount agent $i$ trades at time step $t$ to be proportional to her capital
at that time, denoted by $c_i(t)\geq 0$. More specifically we will
assume that agent $i$ trades a constant fraction $\varepsilon\in [0,1]$ of her
capital. The total bid or so-called excess demand $A(t)$ and the total trading
volume $V(t)$ at time $t$ are therefore given by \be\label{eq:totbid}
A(t)=\sum_{i=1}^{N} \varepsilon c_i(t)
a_i^{\mu(t)},~~~V(t)=\sum_{i=1}^{N} \varepsilon c_i(t),  \ee
where $\mu(t)$ is the piece of information presented to the agents at $t$. 
From (\ref{eq:totbid}) a return $r(t)$ and price process of the underlying
asset and can be defined as follows \cite{dyncap}: 

\be r(t)=\frac{A(t)}{V(t)}, ~~~ \log
p(t+1)=\log p(t) +\frac{A(t)}{V(t)}, 
\ee 

$p(t)$ is here the price of the asset at time $t$.

In a minority game setup the gain for agent $i$ can be written as
\be
g_i(t)=-a_i^{\mu(t)}(t)\varepsilon c_i(t)\frac{A(t)}{V(t)},
\ee
where we follow \cite{dyncap}. Other definitions have been proposed for example in \cite{other,other2}.

We now split the population of $N$ agents into two types, $N_s$ so-called
speculators which we will label by $i=1,\dots,N_s$, and $N_p=N-N_s$ producers,
labelled by $i=N_s+1,\dots,N$.

The capitals of the speculators will be assumed to evolve in time
according to their trading success, 
\BE
c_i(t+1)&=&c_i(t)+g_i(t)\nonumber\\
&=&c_i(t)\left[1-\varepsilon a_i^{\mu(t)}\frac{A(t)}{V(t)}\right]\label{eq:onl},
\EE
where we again follow \cite{dyncap}.

Producers are taken to be agents whose capitals are
drawn from a fixed distribution at the start of the game, and which then do
not evolve in time. The group of producers thus provides a background signal on which the
speculators operate and try to make profit. 

The evolution of the capitals of the speculators can then be written as
follows 
\BE 
&c_i(t+1)=c_i(t)\left[1-\frac{\varepsilon}{V(t)}\sum_j
a_i^{\mu(t)}a_j^{\mu(t)}\varepsilon c_j(t)\right], \quad i=1,\dots,N_s &\label{eq:online}\\
&A(t)=\varepsilon\sum_{i=1}^N
c_i(t)a_i^{\mu(t)}~~~~V(t)=\varepsilon\sum_{i=1}^N c_i(t)&
\label{eq:online2}\EE 
The capitals of the producers $i=N_s+1,\dots,N$ remain constant in time.  

Finally, following the standard MG conventions we take the number $P$ of
different values of the external information $\mu(t)$ to be proportional to
$N$, the total number of players. The ratio $\alpha=P/N$ will then be the key
control parameter of the model, along with the fraction $p=N_s/(N_s+N_p)$ of
producers among the total population of $N=N_s+N_p$ agents.

The key observables we will be interested in in the following are the
average capital of the agents and the so-called predictability of the
market. We will write $\avg{\dots}_{spec}$ and $\avg{\dots}_{prod}$
for averages over the speculators and producers in the following,
i.e. for example
\be
\avg{c}_{spec}=N_s^{-1}\sum_{i=1}^{N_s} c_i,
~~~\avg{c}_{prod}=N_p^{-1}\sum_{i=N_s+1}^{N} c_i
\ee
for the average capitals of the speculators and producers
respectively. Since the capitals of the producers do not evolve in
time, the latter quantity depends only on the distribution from which
the producers weights are drawn at the start of the game.

Following the standard MG conventions \cite{book1,book2} the predictability $H$ will be
defined as follows\footnote{In principle an additional normalisation
with respect to the trading volume given $\mu$ might be desirable
here. This would not however change the qualitative behaviour of $H$
as a function of the model parameters, so that we here stick to the
definition (\ref{eq:hh}).}  \be \label{eq:hh}
H=(PN)^{-1}\sum_{\mu=1}^{\alpha N}\avg{A|\mu}^2, \ee where
$\avg{\dots|\mu}$ denotes an time-average conditional on the
occurrence of information pattern $\mu$. A non-vanishing value of $H$
thus indicates an inefficient market, in which the sign of the excess
demand can be predicted statistically given the knowledge of the state
of the world $\mu$. $H=0$ corresponds to a situation where no such
information can be extracted from the time series of price changes
generated by the model.

\subsection{Analytical solution}
\subsubsection{Effective macroscopic theory}
An analytical solution of the model can be obtained using either a
static approach based on replica techniques or secondly by directly
studying the dynamics via generating functionals. Both approaches have
been used extensively in the MG literature \cite{book1,book2}, we here restrict the
discussion to the dynamical approach, and quote only the main
intermediate steps and final results.

The starting point is the so-called batch dynamics of the model
\be\label{eq:batch}
c_i(t+1)-c_i(t)=-\frac{\varepsilon c_i(t)}{v(t)}\left[\frac{1}{N}\sum_{\mu=1}^{\alpha
    N} a_i^\mu a_j^\mu c_j(t)+h(t)\right], i=1,\dots,N_s \ee 
where an
effective average of the on-line process Eq. (\ref{eq:onl}) over all values of the
external information $\mu\in\{1,\dots,\alpha N\}$ has been performed at each time step,
and where time has been re-scaled appropriately (see \cite{book2} for details regarding this batch limit). $h(t)$ is a
perturbation field introduced to generate response functions, and will
be set to zero at the end of the calculation. $v(t)$ in
(\ref{eq:batch}) is the (re-scaled) total trading volume at time $t$
\be v(t)=\frac{1}{N}\sum_{i=1}^N c_i(t). \ee

The final outcome of the generating functional analysis of this batch
process is an effective theory valid in the limit $N\to\infty$ (at
fixed $\alpha=P/N, p=N_p/N$) and characterised by a non-Markovian
single-agent stochastic process of the form 
\BE\label{eq:effagent}
c(t+1)-c(t)=-\frac{\varepsilon c(t)}{v(t)}\bigg[\alpha\sum_{t'\leq
      t}(\id+G)^{-1}_{tt'}c(t')+\sqrt{\alpha}\eta(t)-h(t)\bigg],
\EE from which the dynamical order parameters are to be determined
self-consistently as 
\BE \label{eq:op}
C(t,t')&=&(1-p)\avg{c(t)c(t')}_\star+p{\avg{c^2}}_{prod},\\
G(t,t')&=&(1-p)\frac{\partial
  \avg{c(t)}_\star}{\partial h(t)},\\
 v(t)&=&(1-p)\avg{c(t)}_\star+p{\avg{c}}_{prod}. \EE  Here $\avg{\dots}_\star$ denotes average over the
Gaussian noise $\{\eta(t)\}$, which in turn comes out to be correlated
in time according to the following co-variance matrix
\be\label{eq:cov} \avg{\eta(t)\eta(t')}_\star =
\left[(\id+G)^{-1}C(\id+G^T)^{-1}\right]_{tt'}. \ee Since this matrix
is defined through the above correlation and response functions $C$
and $G$ the set of equations
(\ref{eq:effagent}-\ref{eq:cov}) form a closed, but
implicit characterisation of the dynamics of the model on the
macroscopic level. This effective theory is exact in the thermodynamic
limit in the sense that a combined site and disorder average
in the original problem, $\overline{\avg{\cdots}}_{spec}$ is
equivalent to the average $\avg{\dots}_\star$ in the resulting
effective-agent description.
\subsubsection{Fixed point analysis}
We will now restrict the further analysis to an inspection of the
fixed points of the effective process (\ref{eq:effagent}). At the
fixed point the correlation function becomes flat \be C(t,t')\equiv q,
\ee and we assume that the response function is invariant against
time-translation $G(t,t')=G(t-t')$ in this regime. The static order
parameters will be given by $q$, the asymptotic average capital
$v=\lim_{t\to\infty} v(t)$ and the integrated response
$\chi=\sum_{\tau\geq 0} G(\tau)$. The single-particle noise
$\{\eta(t)\}$ is then assumed to become static as well at the fixed
point so that $\lim_{t\to\infty}\eta(t)=\eta$ with $\eta$ a time-independent Gaussian
random variable of zero mean and variance $q/(1+\chi)^2$. For
simplicity we will write $\eta=\frac{\sqrt{q}}{1+\chi}z$
in the following, with $z$ a standard Gaussian. Taking into account that all fixed points of
(\ref{eq:effagent}) are to be non-negative, i.e $c\geq 0$, we find 
\be c(z)=-\sqrt{\frac{q}{\alpha}}z\Theta[-z], \ee
with $\Theta[x]$ the step-function, $\Theta[x]=1$ for $x>0$ and
$\Theta[x]=0$ otherwise. The perturbation field $h(t)$ has been set to zero, as announced above. In particular we note that precisely half of the speculators will go bankrupt asymptotically (i.e. their capitals decay to zero), whereas the other half survives.

Recalling $p=N_{p}/(N_{s}+N_{p})$) self-consistency demands 
\BE
v&=&(1-p)\avg{c}_{*}+p{\avg{c}}_{prod},\\
q&=&(1-p)\avg{c^2}_{*}+p{\avg{c^2}}_{prod} \\
\chi&=&-(1-p)\frac{1+\chi}{\sqrt{\alpha q}}\bra\frac{\partial
  c}{\partial z}\ket_{*}, \EE 
from which one finds 
\BE
q&=&\frac{p{\avg{c^2}}_{prod}}{1-\frac{1-p}{2\alpha}},\\
v&=&(1-p)\sqrt{\frac{q}{2\pi\alpha}}+p{\avg{c}}_{prod}, \\
\chi&=&\frac{1-p}{2\alpha+p-1} \label{eq:chi}.\EE
These equations fully characterise the statistics of the assumed fixed point, given the model parameters $\alpha$ and $p$ and the first two moments ${\avg{c}}_{prod},{\avg{c^2}}_{prod}$ of the capitals of producers. From these solutions, the predictability $H$ at the fixed point is then given by
\be 
H=\frac{q}{(1+\chi)^2}, 
\ee
for a derivation of analogous expressions in conventional MGs we refer to \cite{book1,book2}. With a suitable re-scaling $H$ and $v$ become independent of the distribution of producer wealth, in particular one has
\BE
v'&\equiv&\frac{\avg{c}_{spec}}{\sqrt{\avg{c^2}_{prod}}}=\sqrt{\frac{p}{2\pi\alpha\left(1-\frac{1-p}{2\alpha}\right)}}, \\
H'&\equiv&\frac{H}{\sqrt{\varepsilon^2\avg{c^2}_{prod}}}=\frac{p}{\left(1-\frac{1-p}{2\alpha}\right)(1+\chi)^2},
\EE
with $\chi$ as given in (\ref{eq:chi}).

\subsection{Phase transition and stationary states}
\begin{figure}
\vspace{2em}
\centerline{\includegraphics[width=20pc]{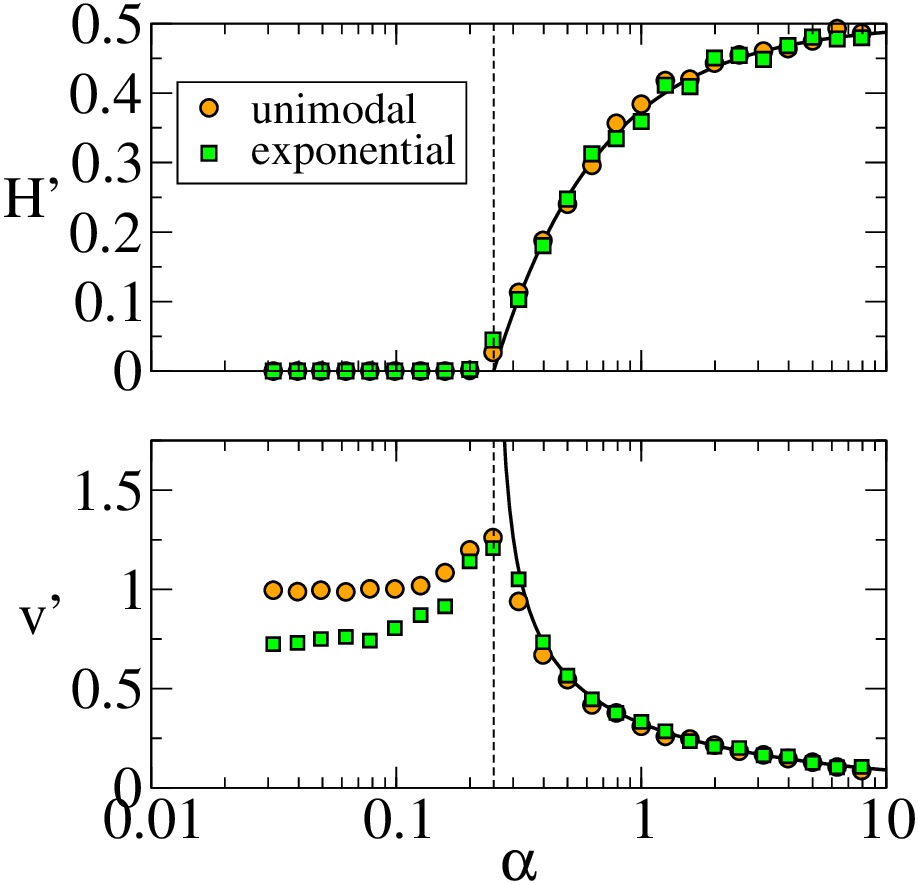}}
\vspace{2em}
\caption{ (colour on-line) Re-scaled wealth
  $v'=\avg{c}_{spec}/\sqrt{\avg{c^2}_{prod}}$ of the speculators and
  re-scaled predictability $H'=H/(\varepsilon^2\avg{c^2}_{prod})$
  versus $\alpha$ for fixed $p=0.5$, and two different distributions
  of the wealth of producers.  Solid lines are the theoretical
  predictions for the phase above the transition at $\alpha_c=0.25$
  (indicated as vertical dashed line).  Symbols are from simulations
  of the on-line process; circles correspond to producers with
  unimodal capital $c_i\equiv 1$, squares to an exponential
  distribution of the producers' capitals.  Simulations are for
  $\alpha N^2=10000$, run for $10^6$ steps (with measurements only in
  the second half of this period).  Averages over $20$ samples of the
  disorder are taken.}
\label{fig:orderparameters}
\end{figure}
We will now briefly discuss the statistical properties of the
stationary states of the model and its phase diagram. Our analytical
predictions for the order parameters of the model in it ergodic
stationary states are verified by numerical simulations in
Fig. \ref{fig:orderparameters}.

Firstly one notes that the phase transition observed in the standard
versions of the MG \cite{book1,book2} also occurs in the present model. In particular we observe a phase with vanishing predictability at low $\alpha$, and a second phase with $H>0$ at values of $\alpha$ above the transition point. Our theory
predicts the integrated response $\chi$ to diverge at a critical value
of $\alpha$ \be \alpha_c=\frac{1-p}{2}, \ee at which the
predictability $H$ vanishes. Below this point the system remains in a
state of zero predictability and divergent integrated response so that
the predictions of the ergodic theory are no longer expected to be
valid.

The occurrence this transition between a phase in which the market is
predictable $H>0$ and a regime in which $H=0$ may be interpreted
geometrically as follows: any stationary solution
$\{c_i\}_{i=1,\dots,N_s}$ of the dynamics which renders
$H=(\alpha N^2)^{-1}\sum_\mu \left(\sum_{i=1}^N a_i^\mu c_i\right)^2$ to
zero necessarily fulfills the $P$ conditions \be \sum_{i=1}^{N_s}
a_i^\mu c_i=-\sum_{i=N_s+1}^N a_i^\mu c_i \ee for
$\mu=1,\dots,P=\alpha N$. The values on right-hand-sides are quenched
random variables, as the only dynamical degrees of freedom are
given by the capitals of the speculators. Within the ergodic theory
$N_s/2$ of the variables $c_i, i=1,\dots,N_s$ are equal to zero as
seen above, so that this defines a linear set of $\alpha N$ equations
for $N_s/2=(1-p)N/2$ unknowns. For $\alpha>(1-p)/2$ the number of
conditions thus outnumber the degrees of freedom so that no solution
with $H=0$ can be found in this regime. Only as
$\alpha=\alpha_c=(1-p)/2$ does the number of conditions become small
enough to allow for solutions with vanishing predictability $H$. Below
$\alpha_c$ such solutions will generally not be unique, hence the
breaking of ergodicity as indicated by the divergence of the
integrated response.

Some other observations can be made at this point

\begin{enumerate}
\item[(i)] the numerical values of the order parameters $v$ and $H$
  depend only on fraction of producers $p$ and show a simple
  dependency on the first two moments of wealth of the producers,
  apart from which the distribution of the weights of the producers is
  irrelevant. The location of the phase transition in turn only
  depends on the relative proportions of producers and speculators.
  This is consistent with \cite{dyncap} where no dependence of the critical
  point on the functional form of the distribution of producer wealth
  and only small qualitative variation of the order parameters was
  reported for the model with dynamical capitals and two trading
  strategies per agent.
\item [(ii)] the analytical theory for the infinite system predicts
  the average capital of the speculators
  $\avg{c}_{spec}=(v-p\avg{c}_{prod})/(1-p)$ to diverge at the phase transition.
  In numerical simulations of finite-size systems we thus expect a
  rounded maximum, as confirmed in Fig. \ref{fig:orderparameters}. We note that very similar
  behaviour is found in the model with two trading strategies per
  player \cite{dyncap}.
\item [(iii)] Inspection of Eq. (\ref{eq:batch}) reveals that the role of $\varepsilon$ in the batch dynamics is essentially to set the overall time scale. In simulations $\varepsilon$ needs to be sufficiently small though to ensure that all capitals remain positive at all times. While the deterministic batch dynamics attains a fixed point in the long-run, this may not necessarily be true for the on-line game. Here dynamical fluctuations are to be expected due to the inherent stochasticity induced by the randomly chosen information patterns $\mu(t)$. The magnitude of these dynamical fluctuations in the agents' capitals is here governed by $\varepsilon$.
\end{enumerate}
\subsection{Stylised facts}
\begin{figure}[t]
\vspace{2em}
\centerline{\includegraphics[width=20pc]{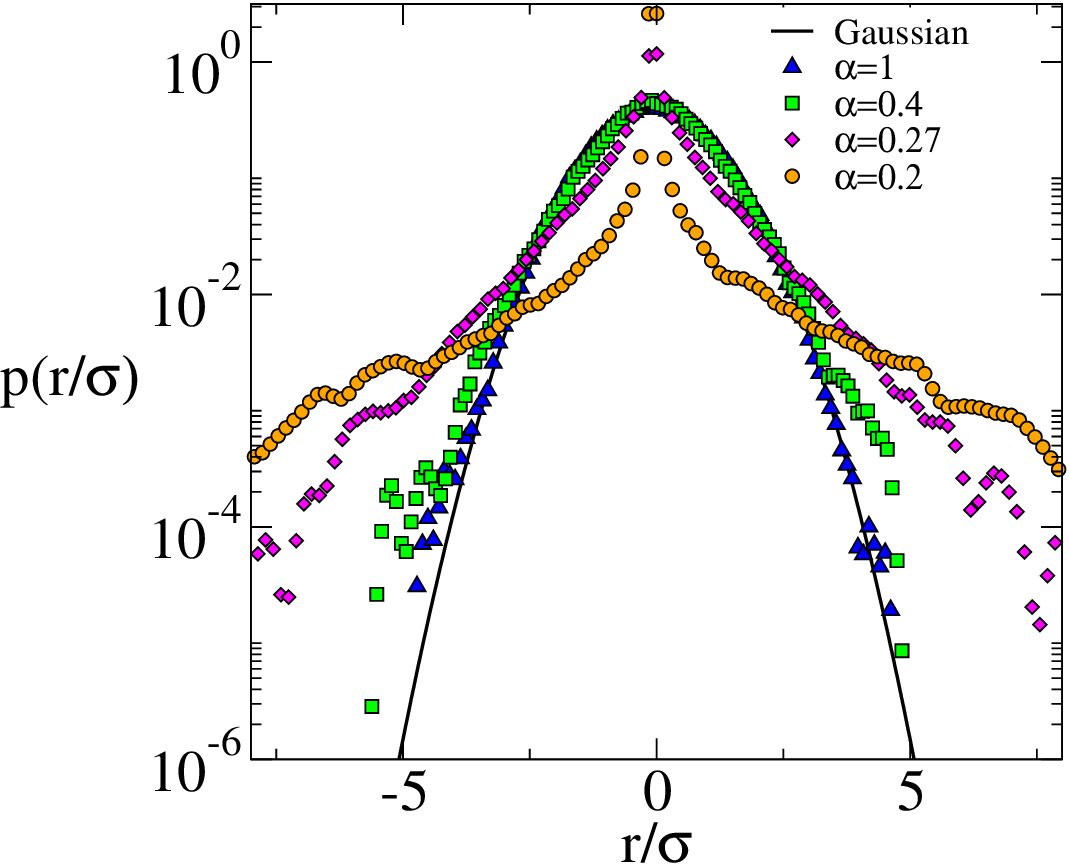}}
\vspace{2em}
\caption{(colour online) Probability distribution of returns $r(t)=A(t)/V(t)$, normalised to
  unit variance. Data is shown for $\alpha=1.,0.4,0.27,0.2$ at
  $p=0.5$ (for which $\alpha_c=0.25$). All simulations are for $N=100$
  agents, run for $10^6$ (on-line) steps, measurements taken in the second half
  of this interval. Data is from $1000$ samples of the disorder. The data at $\alpha=0.2$ contains transients.
}
\label{fig:histogram}
\end{figure}

We now turn to a brief discussion of the statistics of the price time
series generated by the model. In \cite{dyncap} it has been reported that an
MG with dynamical capitals and two trading strategies per player leads
to non-Gaussian price-return distributions close to criticality. This
is confirmed in the present simpler model with only one trading
strategy per player, as shown in Fig. \ref{fig:histogram}. While the
distribution of returns is essentially Gaussian far above the phase
transition, extended tails of large price changes similar to those
observed in real market data develop as $\alpha_c$ is approached from
above. In analogy to GCMGs in which stylised facts persists throughout
an entire critical region, these non-Gaussian effects are observed
throughout the efficient phase of the model at $\alpha<\alpha_c$. We would here like to stress that the examination of the emergence of stylised facts in the simplified MG with evolving capitals is not the focus of this paper. We have therefore not studied the auto-correlation of price returns, or the question as to whether the above non-Gaussian return distributions are indeed a finite-size or transient effect. Further insight might here also be expected from a study of the on-line dynamics with evolving capitals \cite{matteoandi}.
\section{Majority Game with dynamical capitals}
A majority game with evolving capitals can be defined via the following process
\be\label{eq:maj}
c_i(t+1)-c_i(t)=\frac{\varepsilon c_i(t)}{v(t)}\frac{1}{N}\sum_{\mu=1}^{\alpha
    N} a_i^\mu a_j^\mu c_j(t), ~~~~~ i=1,\dots,N, \ee
where the only difference compared to the above minority game is the sign of the interaction term. There is no need to introduce producers here, in fact the majority game is a positive-sum game by definition, and simulating the dynamics of (\ref{eq:maj}) unsurprisingly does not produce much excitement. What is observed as that the total capital grows in time without bound, and simulations soon reach their limits and need to be abandoned after a relatively small number of time steps. One should note that this does not imply that the capitals of {\em all} traders grow to infinity, asymptotically, but rather we find that some capitals diverge, whereas others appear to tend to zero asymptotically \footnote{Due to the discreteness of the dynamics at finite $\varepsilon$, some capitals will become negative eventually, and simulations are aborted at this point.}.

\begin{figure}
\vspace{3em}
\centerline{\includegraphics[width=25pc]{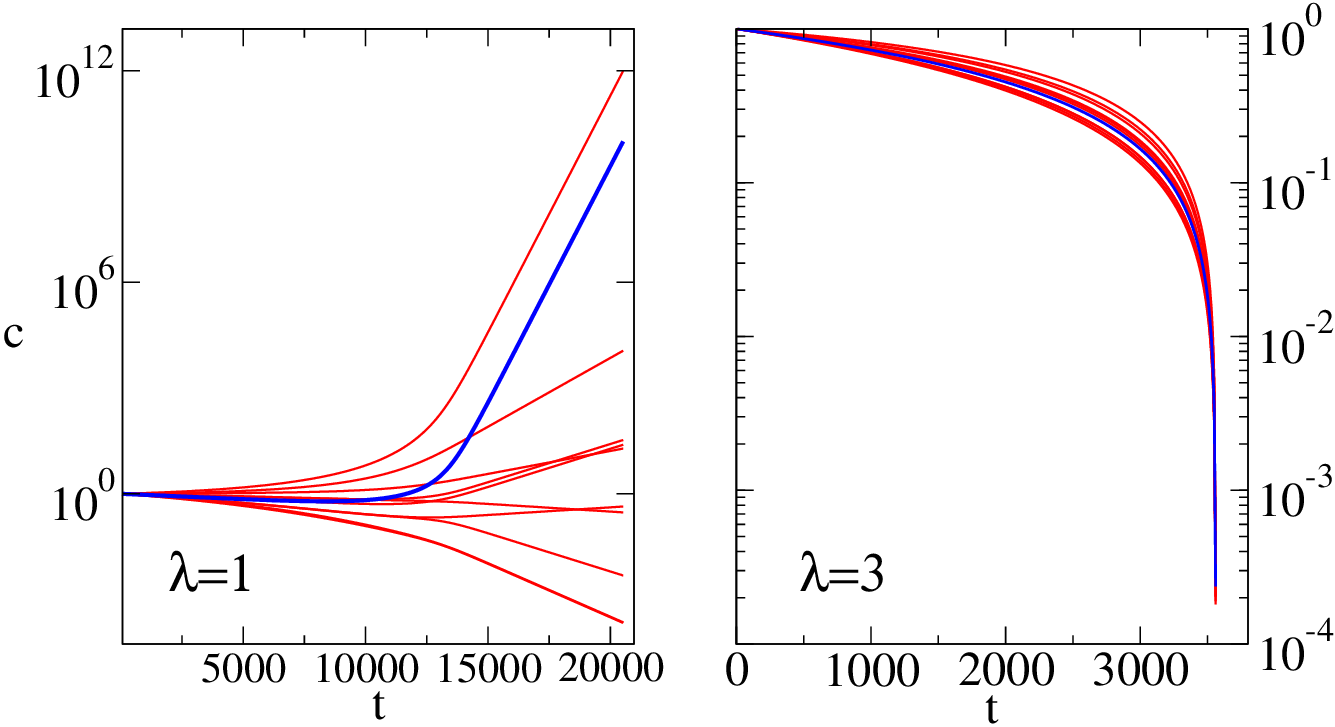}}
\vspace{2em}
\caption{ (Colour on-line) Behaviour of the majority game with evolving capitals, and external forcing term ($\alpha=0.316$, $\varepsilon=10^{-4}$, $N=100$). We show results from single simulation runs, the red (light) curves show the evolution of the capitals of $10$ randomly selected agents, the blue (dark) curves the average capital. In the left panel the forcing parameter is low ($\lambda=1$), so that the average capital grows without bounds. In the right panel the forcing parameter $\lambda=3$ is sufficiently strong to suppress the growth of capitals, and the system approaches a trivial state with all capitals vanishing.}
\label{fig:dyncap_maj}
\end{figure}

Introducing a forcing term, parametrized by $\lambda>0$ as follows
\be
c_i(t+1)-c_i(t)=\frac{\varepsilon c_i(t)}{v(t)}\left[\frac{1}{N}\sum_{\mu=1}^{\alpha
    N} a_i^\mu a_j^\mu c_j(t)-\lambda\right]
\ee
yields a more interesting behaviour, see Fig. \ref{fig:dyncap_maj}. At small, but non-zero values of $\lambda$ the behaviour is as before, the total capital in the system diverges in time. If a large enough value of the limiting force is chosen, however, then simulations suggest an overall contraction of the total capital, i.e. $\lim_{t\to\infty} v(t)=0$. This suggests, that a non-trivial stationary regime may be possible, if $\lambda$ is chosen appropriately to balance the forces between asymptotically diverging, and asymptotically vanishing total capital in the system. This then directly leads to so-called replicator dynamics, used frequently in evolutionary game theory and in population dynamics. We will therefore discuss the behaviour of these equations in the context of the majority game with evolving capitals, subject to an overall conservation, in the next section.
\section{Replicator dynamics}
\subsection{Model definitions}
Replicator equations are a common mathematical framework in evolutionary game theory \cite{Book4}, and generally describe the dynamics of populations of agents, in which each individual, carries one single (pure) strategy. We will label strategies by $i=1,\dots,N$ in the following. The interaction with the other players then leads to a fitness or payoff $f_i$ for strategy $i$. Here a well-mixed interaction with all other members of the population is assumed. If $x_i$ is denotes the proportion of individuals carrying strategy $i$ in the population, then the replicator dynamics are given by
\be
\dot x_i(t)=x_i(t)(f_i[\bx(t)]-f[\bx(t))]).
\ee
We have here explicitly included the frequency-dependence of $f_i$ (i.e. $f_i$ depends on the entire composition $\bx=(x_1,\dots,x_N)$ of the population). $f[\bx(t)]$ is the average fitness in the population, i.e. $f[\bx]=\sum_i x_i f_i[\bx]$. Provided the concentration vector $\bx$ is normalised when the dynamics are started, i.e. $\sum_i x_i(t_0)=1$, this normalisation will be maintained at all later times. Further details on replicator dynamics in the context of evolutionary game theory can be found for example in \cite{Book4}. It is now not difficult to see the analogy with the dynamics discussed above in the context of the majority game with evolving capitals. The $\{c_i\}$ play the role of the concentrations $\{x_i\}$, for convenience and consistency with the replicator literature, we will use $x_i$ from now on. The fitness of a given strategy $\ba_i$ in a population of composition $\bx$ is then given by
\be
f_i=\frac{1}{N}\sum_{j,\mu} a_i^\mu a_j^\mu x_j
\ee
in the context of the majority game. Introducing
\be
J_{ij}=\frac{1}{N}\sum_{\mu=1}^{\alpha N} a_i^\mu a_j^\mu
\ee
and the using the convention of vanishing diagonal elements $J_{ii}=0$, we then use
\be\label{eq:repl}
\frac{d}{dt}x_i(t)=x_i(t)\left[-2u x_i(t)+\sum_j J_{ij}x_j(t)-\lambda(t)\right].
\ee
as a starting point for our analysis. Note here that the strategy concentrations have been re-scaled and normalised to $N^{-1}\sum_i x_i(t)=1$. This guarantees a well-defined thermodynamic limit. The first term in the square brackets has been introduced following \cite{Book5,DO,OD,OD2,myhebb}, and describes what is referred to as `co-operation pressure'. It has been seen in the context of \cite{OD,myhebb,myasym} that large positive $u$ drive replicator systems towards the centre of the strategy simplex, and one expects a stable fixed point at which all strategies are played with equal frequency. For low values of $u$ the dynamics can approach the boundary of the simplex, and a sizable fraction of strategies is not used at all asymptotically. While we leave $u$ as a general model parameter, it is worth pointing out that for the choice $u=-\alpha/2$, Eq. (\ref{eq:repl}) is precisely the replicator dynamics corresponding to the majority game. $\lambda(t)$ is again the average fitness (now given by $\lambda=N^{-1}\sum_{ij}J_{ij}x_i x_j$), and can be treated as a general Lagrange multiplier maintaining the overall normalisation \footnote{Strictly speaking the normalisation is only conserved as a soft constraint in our analytical calculation, i.e. $N^{-1}\sum_i x_i(t)=1$ holds as an average over the disorder. In our simulations the constraint is respected for each sample.}.
\subsection{Analytical solution}
The model defined above is closely related to that studied in \cite{myhebb}, but with an opposite overall sign in the Hebbian-type dynamics. As we will see the resulting phase diagram is different from that of the model of \cite{myhebb} however. Analytical progress can be made using again the formalism of dynamical mean-field theory. Similar to \cite{OD,myhebb} we find the following effective single-species process:
\BE
\hspace{-5em}\frac{d}{dt}x(t)&=&x(t)\bigg[-2u x(t)-\alpha x(t)+\alpha\int dt' (\id+G)^{-1}(t,t')x(t')+\sqrt{\alpha}\eta(t)-\lambda(t)\bigg],
\EE
where
\be
\avg{\eta(t)\eta(t')}_\star =
\left[(\id+G)^{-1}C(\id+G^T)^{-1}\right]_{tt'}, \ee
and the corresponding self-consistency relations
\be
C(t,t')=\avg{x(t)x(t')}_*, ~~~ G(t,t')=\avg{\frac{\delta x(t)}{\delta\lambda(t')}}_*, ~~~ \avg{x(t)}_*=1
\ee
Following \cite{OD,myhebb} we will next look for fixed points of this effective stochastic process. These fulfill  
\BE
x\bigg[-2u x-\alpha\frac{\chi}{1+\chi}x+\sqrt{\alpha}\eta-\lambda(t)\bigg]=0,
\EE
i.e. one has $x=0$ or
\BE
x=\frac{\sqrt{\alpha}\eta-\lambda(t)}{2u+\alpha\frac{\chi}{1+\chi}}
\EE
We will write $\eta=-\frac{\sqrt{\alpha q}}{1+\chi}z$ with $z$ a standard Gaussian, and similar to what was proposed first in \cite{OD} for a replicator model with Gaussian couplings we use the physical ansatz
\be
x(z)=\frac{\frac{-\sqrt{\alpha q}}{1+\chi}z-\lambda}{2u+\alpha\frac{\chi}{1+\chi}}\Theta\left[-\frac{\sqrt{\alpha q}}{1+\chi}z-\lambda\right]
\ee
to proceed, i.e. one has
\be
x(z)=\frac{\sqrt{\alpha q}}{1+\chi}\frac{\Delta-z}{2u+\alpha\frac{\chi}{1+\chi}}\Theta\left[\Delta-z\right]
\ee
where $\Delta=-\lambda/\frac{\sqrt{\alpha q}}{1+\chi}$. From this, following the lines of \cite{OD,myhebb} one derives the following self-consistent equations for the static order parameters $\chi$ (the integrated response as before), $q=\avg{x^2}_*$ and the asymptotic disorder-averaged Lagrange multiplier $\lambda$

\BE
-\left[2u+\alpha\frac{\chi}{1+\chi}\right]\chi&=&f_0(\Delta) \label{eq:r1}\\
\frac{1+\chi}{\sqrt{\alpha q}}\left[2u+\alpha\frac{\chi}{1+\chi}\right]&=&f_1(\Delta)\label{eq:r2} \\
\frac{(1+\chi)^2}{\alpha}\left[2u+\alpha\frac{\chi}{1+\chi}\right]^2&=&f_2(\Delta)\label{eq:r3} 
\EE
with $f_n(\Delta)=(2\pi)^{-1/2}\int_{-\infty}^\Delta dz~ e^{-z^2/2}(\Delta-z)^n$. In the fixed-point regime $q$ and $\lambda$ are time-independent, and $\chi$ is a stationary order parameter by definition. Using (\ref{eq:r1},\ref{eq:r2},\ref{eq:r3}) these order parameters can be obtained numerically as functions of the model parameters $u$ and $\alpha$.

Before we present results in the next section, it is worth investigating the stability of the fixed-points which we have assumed to derive the above closed equations describing the stationary state. The stability can here be determined within a linear expansion about the assumed fixed-point. This is performed on the level of the effective process and follows the lines of \cite{OD,OD2}. We will not present the details of the calculation here, a result similar to \cite{myhebb,OD} is found: at large values of $u$ greater than a critical value $u_c(\alpha)$, the above solutions is indeed stable and valid. Below $u_c$ the system may still approach a fixed point asymptotically, but generally the dynamics allow for a large number of attractors and none of the fixed points is locally stable, as perturbations do generally not decay \cite{DO}. As in \cite{OD,OD2}, for each $\alpha$ the transition point $u_c$ coincides with the point at which $\phi=1/2$, i.e. one here has $\Delta=0$. Using this condition, $u_c$ is then determined numerically from (\ref{eq:r1},\ref{eq:r2},\ref{eq:r3}) (solving these equations delivers $q=1/\pi$ at $u=u_c$ along with a value for $\chi$ at the transition, which we do not report here). 
\subsection{Stationary state and phase diagram}
\begin{figure}
\vspace{2em}
\centerline{\includegraphics[width=20pc]{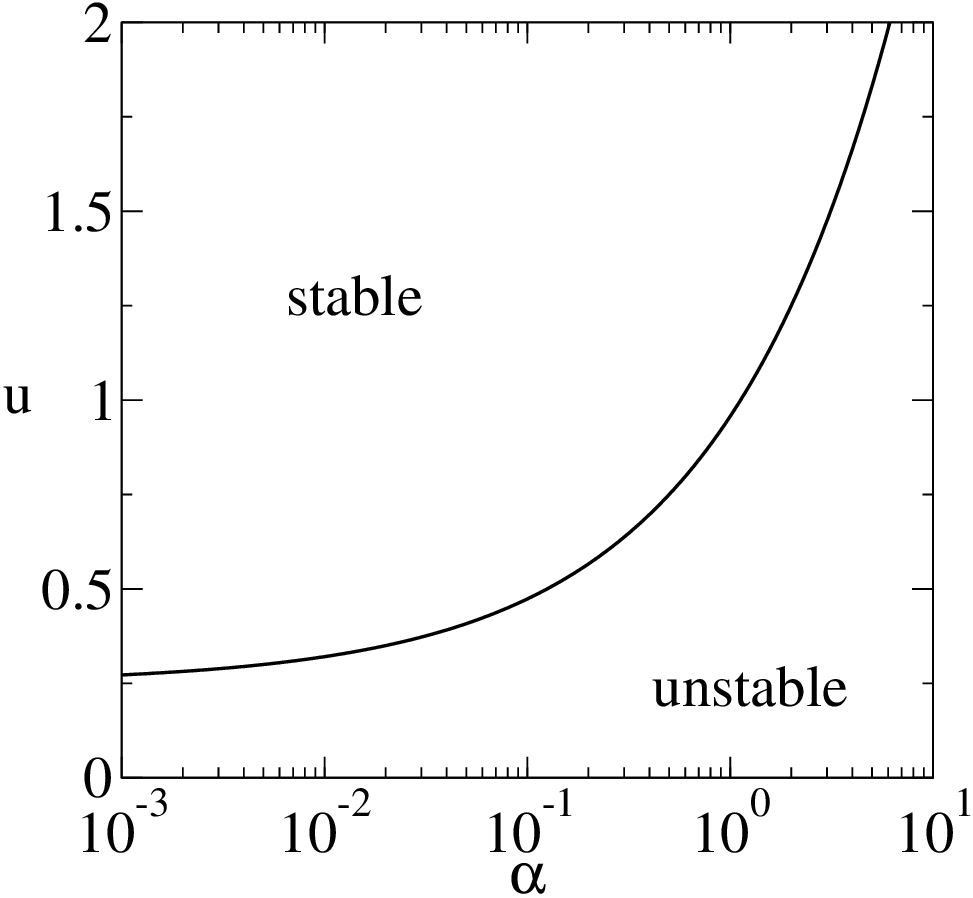}}
\caption{Phase diagram of the replicator model. The line separating the stable and unstable phases is obtained within a linear stability analysis. At the transition precisely half of the initially present strategies carry non-zero weight asymptotically, i.e. one has $\phi=1/2$. In the unstable phase the system may still approach fixed-points asymptotically, but the number of different attractors is exponential in the system size (see also \cite{DO,OD,OD2} for related systems).}
\label{fig:pg_antihebb}
\end{figure}
The resulting phase diagram is shown in Fig. \ref{fig:pg_antihebb}. The line shows the onset of instability in the $(\alpha,u)$ plane. It is here also worth noting that (\ref{eq:r1},\ref{eq:r2},\ref{eq:r3}) do not allow for a divergence of the susceptibility $\chi$. This is prevented for example by Eq. (\ref{eq:r1}): given that $f_0(\Delta)$ is finite and bounded in the interval $[0,1]$ for all $\Delta$, assuming a divergence of $\chi$, i.e. $|\chi|\to\infty$ directly leads to $u=-\alpha/2<0$. Given that an instability of the fixed point occurs at a positive value $u_c(\alpha)$ when the co-operation pressure is lowered from large positive values, such a divergence of $\chi$ never occurs in the physical system \footnote{We here note that the transformation $u\to-u$ turns the present model into that discussed in \cite{myhebb}, but with an overall minus sign in front of the right-hand side of the replicator dynamics (corresponding to $t\to-t$). In \cite{myhebb} a transition with a divergence of the integrated response was found at $\alpha=u/2$, in particular this transitions occurs when $u$ is lowered from a stable fixed point state at $u>\alpha/2$. Due to the inversion in signs, stability properties are also inverted in comparison to the model discussed here, so that what used to be a stable fixed point regime is now unstable, and no such transition is seen in the present model.}  

\begin{figure}
\vspace{4em}
\centerline{\includegraphics[width=20pc]{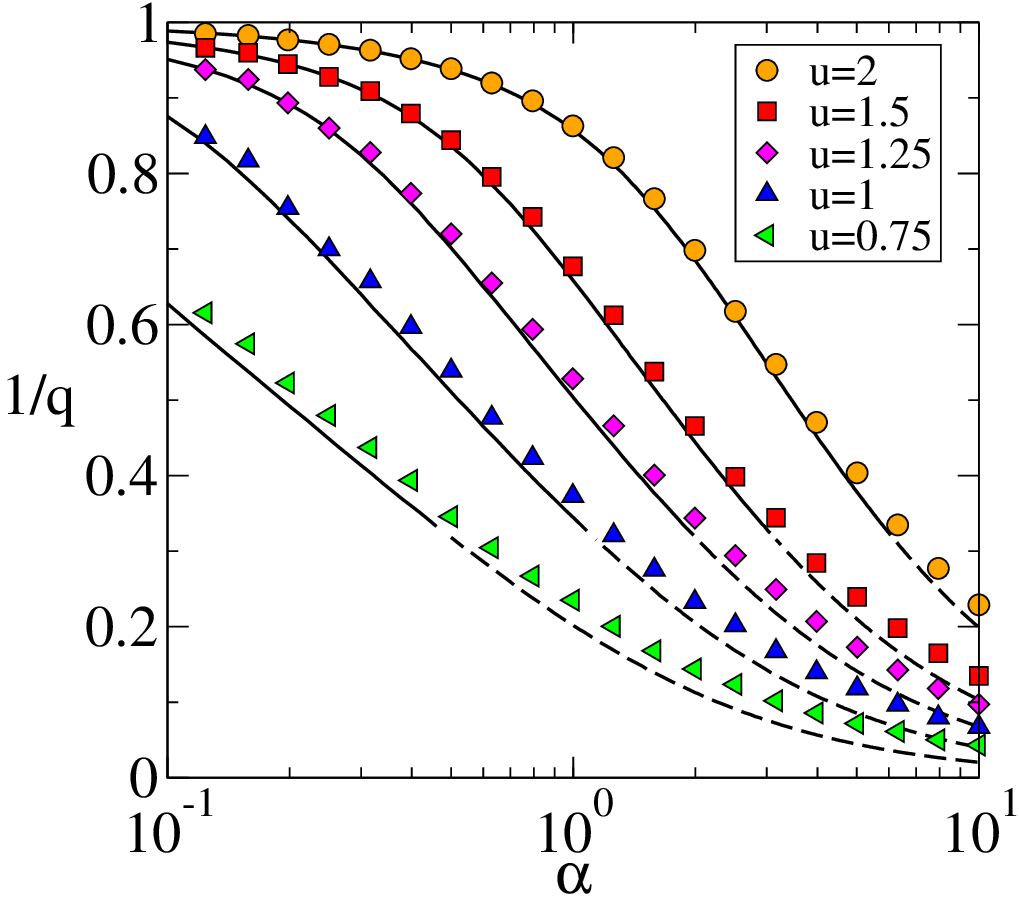}}
\vspace{2em}
\caption{(Colour on-line) Order parameter $1/q$ as a function of $\alpha$ for the replicator model at different values of the co-operation pressure. Simulations (markers) are performed for $N=500$ agents, averaged over $50$ samples, and using a discretisation of the replicator equations as proposed in \cite{OD2}. Solid lines are from the theory in the stable phase, continued as dashed lines into the unstable region, where the theory is no longer valid.}
\label{fig:q_antihebb}
\end{figure}

\begin{figure}
\vspace{4em}
\centerline{\includegraphics[width=20pc]{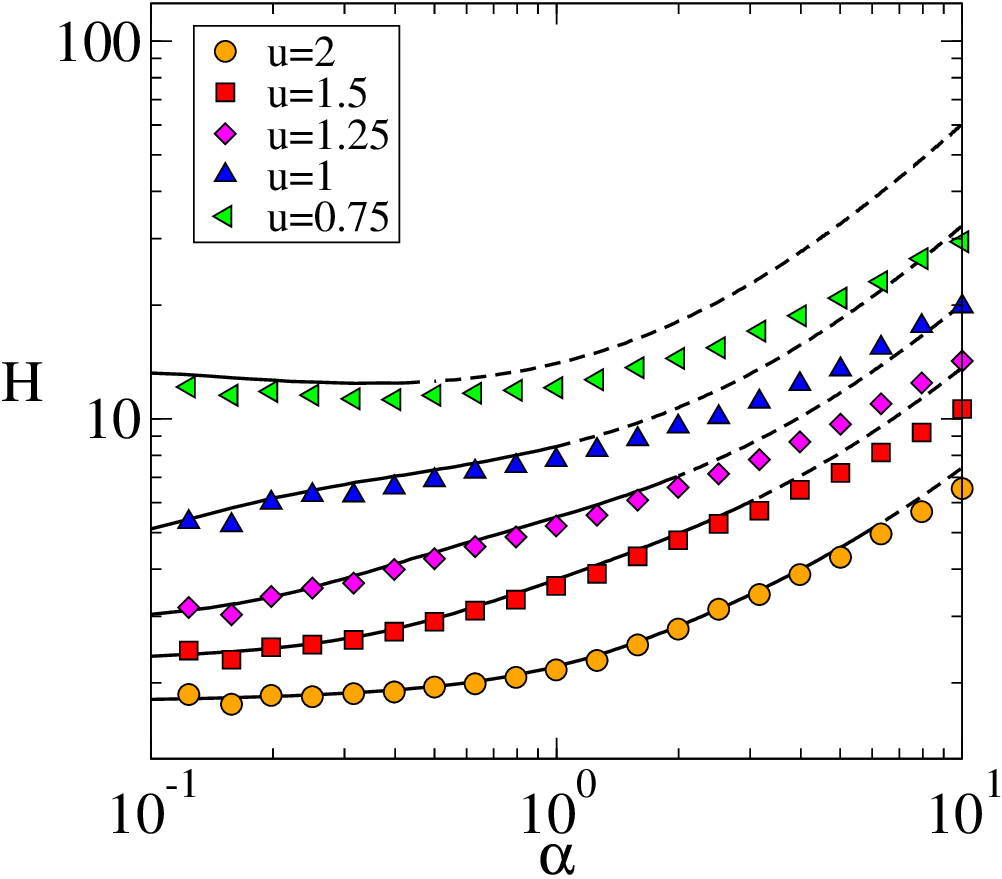}}
\vspace{2em}
\caption{(Colour on-line) The analogue of the predictability $H$ as function of $\alpha$ for the replicator model at different values of the co-operation pressure. Markers show results from simulations (same parameters as in Fig. \ref{fig:q_antihebb}). Solid lines are from the theory in the stable phase, continued as dashed lines into the unstable region, where the theory is no longer valid.}
\label{fig:h_antihebb}
\end{figure}

\begin{figure}[t!]
\vspace{4em}
\centerline{\includegraphics[width=20pc]{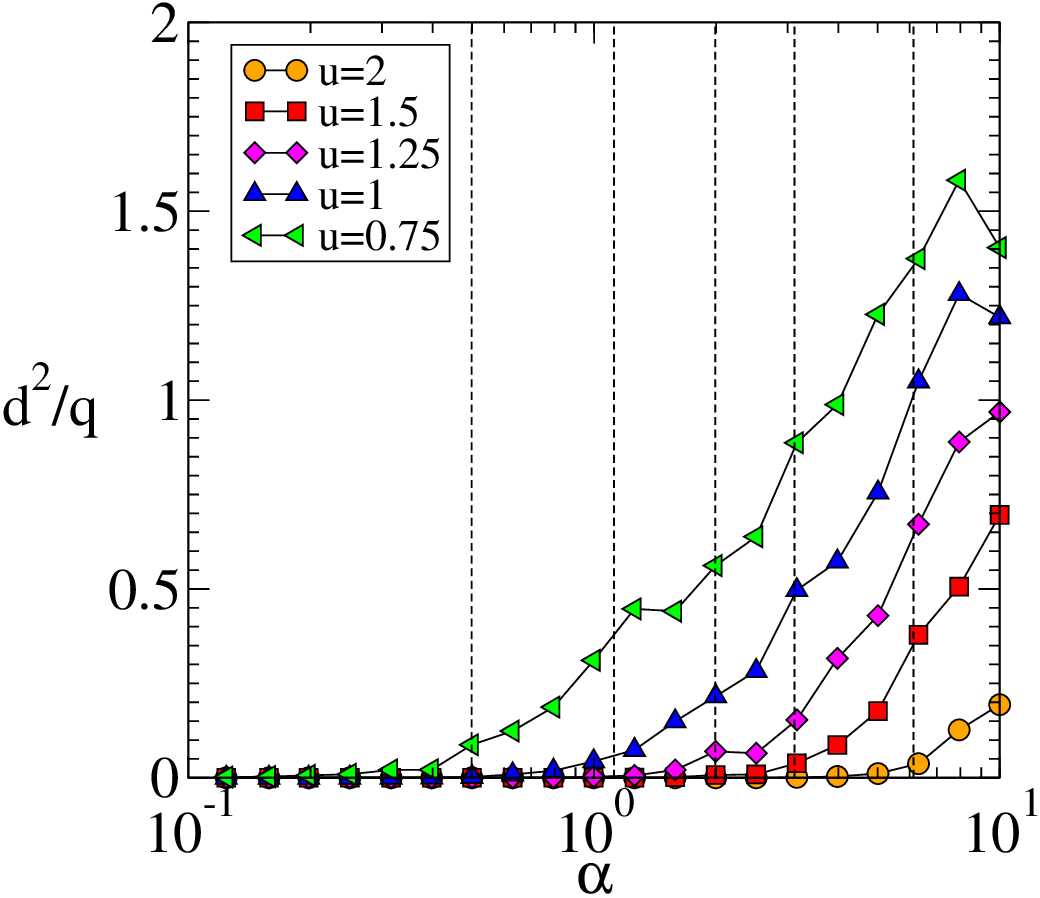}}
\vspace{2em}
\caption{(Colour on-line) Sensitivity to initial conditions as measured by the distance between runs at fixed choices of the random disorder, but started from different initial conditions (see text for further details). Markers show results from simulations (same parameters as in Fig. \ref{fig:q_antihebb}), and are connected for optical convenience. The vertical lines mark the location of the phase transition and onset of instability, as predicted by the theory.}
\label{fig:dist}
\end{figure}

The validity of the generating functional theory is confirmed in Figs. \ref{fig:q_antihebb} and \ref{fig:h_antihebb}, where we show results for the order parameters $1/q$ and $H$ as a function of $\alpha$ for different choices of $u$. At small values of $\alpha$, where the theory is predicted to be stable and accurate, near perfect agreement between theory and simulations is observed. Only at large value of $\alpha$ does the fixed point ansatz become unstable, and we find systematic deviations. Fig. \ref{fig:q_antihebb} here demonstrates that the asymptotic diversity of the system decreases as $\alpha$ increases. To this end note that $1/q$ is similar to what is known as Simpson's index \cite{simpson} in ecology, and a measure of how many distinct species (i.e. strategies) survive in the long run. In the extreme case of only one surviving strategy (with concentration $x=N$) one has $q=N$, so $1/q\to 0$ in the thermodynamic limit. On the other hand, if all species contribute at equal concentration, $x_i=1$ for all $i=1,\dots,N$, then $q=1$. Fig. \ref{fig:h_antihebb} shows that the predictability $H$ remains strictly positive for all studied values of $u$ and $\alpha$, again confirming the absence of a transition with diverging integrated response and of a phase with vanishing predictability and optimal resource exploitation.
In Fig. \ref{fig:dist} we characterise the nature of the instability transition depicted in the phase diagram of Fig. \ref{fig:pg_antihebb} further. To this end we have, for a fixed choice of the disorder (i.e. for fixed strategy assignments $\ba_i$) started the system from two different initial conditions $\bx(t=0)$ and $\bx'(t=0)$, and have then monitored the distance $d^2(t)=\frac{1}{N}\sum_i (x_i(t)-x'_i(t))^2$. The figure shows results at long times, and re-scaled suitably by a factor of $q$. As seen in the diagram, the asymptotic distance vanishes at large values of $u$, greater than the critical $u_c$ predicted by the theory, but remains non-zero below. We conclude that initial conditions are irrelevant in the stable phase, and that the replicator dynamics converges to a single unique fixed point for any realisation of the disorder. Below the transition, multiple attractors are possible, and which of these is chosen is determined by the initial conditions from which the dynamics is started. Transitions of this type have been related to what is called `memory onset' in MGs \cite{impact2} in \cite{myasym, myhebb}.

Finally, one may wonder whether it is possible to study the minority game with dynamical capitals as well in the context of replicator dynamics. This is of course the case, and leads to a replicator model, essentially with an inverted sign in front of the Hebbian interaction term in (\ref{eq:repl}). This model, termed replicators with Hebbian interactions, has been studied in \cite{myhebb}\footnote{In light of the present study it would probably be more appropriate to refer to the model of \cite{myhebb} as replicator dynamics with anti-Hebbian couplings (corresponding to a minority game), and to the present model as one with Hebbian couplings (majority game).}, so that we do not pursue this here, but only mention that both types of transitions (one with diverging integrated response, and one indicated by a continuous onset of memory) are observed in this case.

\section{Discussion and concluding remarks}

In summary our study demonstrates that the minority game with
dynamical capitals put forward in \cite{dyncap} can be simplified to a
model with only one trading strategy per player without losing many of
the features of the more complicated model in which every agents holds
two strategies. In particular the transition between a symmetric phase
at low values number of available information patterns, and an
asymmetric phase at large value of the key model parameter $\alpha$
remains. The theoretical analysis of the simpler model is
straightforward and can be performed without much technical effort
based on the standard methods discussed in the
literature\cite{book1,book2}. As in many other variants of the minority
game the dynamic susceptibility (or integrated response) diverges at
this transition. Simulations in both the simplified and the original
model with evolving capitals reveal a maximum of the average wealth
per speculator at the phase transition point. Our analytical
calculations show that this quantity diverges in the infinite system
at the transition, the maximum observed in simulations is a reflexion of this in finite systems. At the same time the model can lead to non-Gaussian wealth distributions around and below the transition. 

We have then furthermore briefly discussed a majority game with evolving capitals. The dynamics of this model is relatively unspectacular, unless forced by an external decay term, the total capital in this positive-sum game grows without bounds, and diverges asymptotically. Unless a mixture of minority and majority game players is considered (which we have not done here) non-trivial dynamics in the majority game with evolving capitals can only be expected if a suitably chosen Lagrange multiplier is introduced, keeping the total wealth constant in time. This established an immediate connection to replicator dynamics with interactions of the Hebb type. Such models can be studied with the same analytical methods, and the analysis reveals that a different type of transition is observed in such circumstances. While no divergence of the integrated response and no symmetric phase with vanishing predictability is found, a dynamical instability separating a regime with one unique stable fixed point from a second phase with multiple attractors can be identified. Initial conditions are seen to be irrelevant in the stable phase, but determine the choice of attractor below the instability.

To conclude let us mention that future work might be concerned with the characterisations of the differences between on-line and batch minority game models with evolving capitals. The simple model discussed in the first part of this paper might provide a suitable starting point, and first steps in the direction of addressing the corresponding on-line game have been taken in \cite{matteoandi}. Progress can here probably be made using the tools discussed in \cite{book2}, where on-line versions of the minority game have been addressed, even though more basic approaches might suffice in the context of the simple model discussed here. Secondly, one may think of mixed minority and majority games \cite{mixed} with dynamical capitals, and investigate whether these show a phase transition, even if each agent is equipped only with one strategy. Finally, the MG with dynamical capitals and
multiple strategies per player can be expected to be accessible by
dynamical and static methods as well as pointed out in \cite{dyncap},
although the analysis appears here much more intricate due to the
interplay of slowly-evolving capitals and fast spin degrees of freedom
characterising which of their strategies the individual agents chose to
play. While the derivation of a description of the effective dynamics
within the generating functional formalism of dynamical mean field theory seems straightforward (but tedious) the subsequent analysis of the
resulting effective process is likely to involve some type of
adiabatic approximation to disentangle the different time-scales of
the two types of degrees of freedom.

\section*{Acknowledgements} TG is supported through an RCUK Fellowship (RCUK reference EP/E500048/1). Earlier funding by the European Community's Human Potential Programme under contract HPRN-CT-2002-00319, STIPCO, held at ICTP Trieste, is acknowledged. This work was finalised at the Research Centre for Complex Systems Science, University of Shanghai for Science and Technology, where TG was a visiting fellow of the Shanghai Academy of System Science. Its kind hospitality is appreciated. The authors would like to thank Damien Challet and Matteo Marsili for helpful discussions.

\end{document}